  \documentclass[aps,twocolumn,superscriptaddress,showpacs]{revtex4}

\usepackage{latexsym}
\usepackage{graphicx}

\begin{document}

\title{Structure of the superconducting state in a fully
frustrated wire network \\ with dice lattice geometry}

\author{S. E. Korshunov}
\affiliation{L. D. Landau Institute for Theoretical Physics,
Kosygina 2, Moscow 119334, Russia}
\affiliation
{Laboratoire de Physique Th\'{e}orique et Hautes
\'{E}nergies, CNRS UMR 7589,  \\
Universit\'{e} Paris VI and VII,
4 place Jussieu, 75252 Paris Cedex 05, France}

\author{B. Dou\c{c}ot}
\affiliation{Laboratoire de Physique Th\'{e}orique et Hautes
\'{E}nergies, CNRS UMR 7589,  \\
Universit\'{e} Paris VI and VII,
4 place Jussieu, 75252 Paris Cedex 05, France}

\date{October 22, 2004}

\begin{abstract}
The superconducting state in a fully frustrated wire network
with the dice lattice geometry is investigated in the vicinity
of the transition temperature.
We express the projection of the Ginzburg-Landau free energy
functional on its unstable subspace
in terms of variables
defined on the triangular sublattice of sixfold coordinated sites.
For the resulting effective model, we construct a large class of
degenerate equilibrium configurations, which are in one to one
correspondence with ground states of the fully frustrated $XY$ model
with a dice lattice. The entropy of this set of states is proportional
to the linear size of the system. Finally, we show that
magnetic interactions between currents provide a degeneracy
lifting mechanism and find the structure of the periodic state
selected by these interactions.
\end{abstract}


\maketitle

\section{introduction}

The concept of frustration has been a common link among various
problems in statistical mechanics for the past two decades at least.
Even in the absence of disorder, it often results in a phenomenon of
competition between several degenerate ground-states. Superconducting
wire networks provide a very appealing class of systems where many subtle
effects induced by frustration can be observed experimentally and
analyzed theoretically~\cite{dG,A,AH,PCR,PCRV,WRP,SL,GG,NN}.

For simple regular networks, a natural parameter characterizing
the strength of the frustration is the ratio $f=\Phi/\Phi_{0}$
where $\Phi$ is the external magnetic flux through an elementary
plaquette of the lattice and $\Phi_{0}=hc/2e$ is the
superconducting flux quantum. For an ideal network of very thin
wires, all physical properties are expected to be periodic
functions of $f$, all integer values being equivalent. In this
case, the maximal frustration is obtained when $f$ reaches
half-integer values. Such fluxes are interesting because already
for a single loop, they provide two equivalent ground states,
distinguished by the orientation of the supercurrent flowing
around the loop. For more complex geometries, two adjacent loops
(sharing a common link) have a lower free energy when the currents
in them flow in opposite directions. The possibility to fulfill
this requirement for any pair of adjacent loops is a geometrical
property of a given lattice, which allows one to be sure about the
structure of the superconducting state without any additional
analysis.  This clearly holds for a square lattice~\cite{TJ}
(where vortices of alternating signs form a regular checkerboard
pattern) or for a triangular lattice.

In recent years, network geometries which do not satisfy this criterion
have received a lot of attention. The most studied examples are
the honeycomb~\cite{Teitel85,Shih85,Korshunov86,Xiao02,Lin02}, the
{\em kagome}~\cite{Xiao02,Lin02,LN,HR,Rzchowski97,Higgins00,PH,K02},
and the dice~\cite{VMD,Abil,Pannetier01,Serret02,K01,Cataudella03}
lattices. On the honeycomb lattice, the discrete degeneracy of the
classical ground states in fully frustrated
superconducting wire networks or Josephson junction arrays
can be described in terms of formation of zero-energy
domain walls in parallel to each other~\cite{Korshunov86}, the residual
entropy of such system being proportional to its linear size. Experimentally,
a cusp-like local maximum in the superconducting transition temperature $T_c$
is observed as the external magnetic field is varied around the value
corresponding to $f=1/2$~\cite{Xiao02}. This behavior has been interpreted
as an evidence for a degeneracy lifting mechanism which selects a
commensurate ordered pattern of vortices~\cite{Xiao02}.

For the {\em kagome} lattice, the residual entropy of classical
ground states is much larger, since it is proportional to the whole
network area~\cite{HR,Elser}. The experimental situation at
$f=1/2$ is not as clearcut as for the honeycomb lattice, since the shape
of $T_c$ versus magnetic field curves near $f=1/2$ depends on the resistive
criterion chosen to determine $T_c$~\cite{Higgins00}, or on the superconducting
metal (aluminium versus niobium for instance~\cite{Xiao02}).
Theoretically, various degeneracy lifting mechanisms have been studied in
detail by Park and Huse~\cite{PH}.

\begin{figure}[b]
\includegraphics[width=5cm]{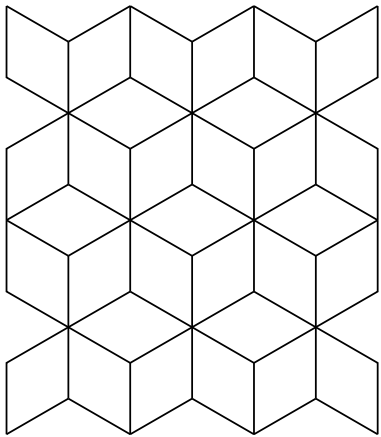}
\caption[Fig. 1]
{Finite cluster with the dice lattice geometry.}
\label{fig1}
\end{figure}

On the dice lattice (see Fig.~\ref{fig1}),
the residual entropy is proportional
to the system linear size~\cite{K01}, as for the honeycomb lattice.
Experimentally, magnetic decoration experiments~\cite{Pannetier01,Serret02}
have found a highly disordered vortex pattern, with a vortex correlation length
comparable to the lattice spacing. Numerical simulations~\cite{Cataudella03}
of the corresponding $XY$ model support the picture proposed in
Ref.~\onlinecite{K01} for the ground states, but also
demonstrate, at low temperatures, an unusually slow relaxation of
energy, as well as aging of phase correlation functions.

In this article we consider maximally frustrated superconducting
networks on a dice lattice, in the immediate vicinity of the
superconducting transition temperature. In this limit, the
amplitude of the superconducting order-parameter is not
necessarily uniform, and it is appropriate to use a generalization
of the approach introduced by Abrikosov~\cite{Abr} in his first
prediction of vortex lattices in type II superconductors. The main
idea underlying this approximation is that at $T_c(H)$, only a
small fraction of the eigenmodes of the linearized Ginzburg-Landau
equations become unstable. Abrikosov developed a variational
procedure where the superconducting order-parameter is constrained
to remain in this unstable subspace (for $T\rightarrow T_c(H)$
this procedure is asymptotically exact). Minimizing the quartic
term in the Ginzburg-Landau free energy functional yields then
periodic vortex lattice solutions.

Adopting this approach to a dice lattice network is quite
interesting since the corresponding eigenmodes (for $f=1/2$) have
unusually high degeneracy and exhibit the unexpected property of
an extreme form of spacial localization. It is indeed possible to
construct an eigenfunction basis for which each member is
non-vanishing only on a {\em finite} cluster~\cite{VMD,Vidal01}.
This phenomenon arises from the Aharonov-Bohm interference effect
which is magnified in the geometry of the dice lattice, and these
{\em Aharonov-Bohm cages} have been evidenced experimentally by
the observation of magnetoresistance oscillations in ballistic
semiconductor networks~\cite{Naud01} with the flux period $hc/e$
per elementary loop.

The main result of the present study is that in maximally
frustrated superconducting wire network non-linear effects select
a class of order-parameter configurations in one to one
correspondence with the ground states of the fully frustrated $XY$
model with the same geometry~\cite{K01}, which may be viewed as a
low temperature approximation for the Ginzburg-Landau model
ignoring amplitude variations (London limit). However, the
inclusion into analysis of the magnetic energy leads to the
removal of the accidental degeneracy and selection of one of the
periodic states minimizing the Ginzburg-Landau free energy.
The same state has the lowest free energy also at lower
temperatures (down to London limit), as well as in Josephson
junction arrays with the same geometry.

In sections~\ref{HarmonicFE} and~\ref{quarticFE},
we express the Ginzburg-Landau functional for a fully frustrated
dice lattice wire network after projection on the subspace of
unstable modes in terms of complex variables defined on the triangular
sublattice of sixfold coordinated sites.
Section~\ref{Minimization} describes the construction
of periodic equilibrium states for this effective problem,
and their extension to a larger class of degenerate states whose
precise connection with those proposed for the corresponding $XY$
model is established.
Finally, section~\ref{Selection} investigates a degeneracy
lifting due to magnetic interaction between currents.

\section{Harmonic contribution to free energy}
\label{HarmonicFE}
\subsection{A single wire}

In the framework of the Ginzburg-Landau approximation the free
energy of a thin superconducting wire, $F^{\rm GL}_{\rm wire}$,
can be written as the sum of the two terms,
\begin{eqnarray}
F^{\rm (2)}_{\rm wire}
&=&\int_{0}^{L}dx\,\left\{-\frac{\alpha}{2}|\Delta(x)|^2 \mbox{}+
\right.\label{a1a}\\
& &\left. \mbox{}+\frac{\gamma}{2}\left|\left[-i\frac{\partial}{\partial x}
-\frac{2\pi}{\Phi_0}A_\|(x)\right]\Delta(x)\right|^2\right\} \nonumber
\end{eqnarray}
and
\begin{equation}
F^{(4)}_{\rm wire}=\frac{\beta}{4}\int_{0}^{L}dx\,|\Delta(x)|^4
\;,                                                 \label{a1b}
\end{equation}
describing, respectively, the harmonic and the fourth-order
contributions to $F^{\rm GL}_{\rm wire}$. Here
\makebox{$\alpha\propto T_{c0}-T$}, $\beta$ and $\gamma$ are the
coefficients of the Ginzburg-Landau expansion, $L$ is the length
and $T_{c0}$ the mean field transition temperature of the wire,
$\Delta(x)$ is the superconducting order parameter as a function
of the coordinate $x$ along the wire, $A_\|(x)$ is the projection
of the vector potential on the wire and $\Phi_0=hc/2e$ is the
superconducting flux quantum.

At the point of phase transition $|\Delta(x)|\rightarrow 0$ and
$F^{(4)}_{\rm wire}$ can be neglected in comparison with
$F^{(2)}_{\rm wire}$. For the given values of $\Delta(x)$ at the
ends of the wire,
\begin{equation}
\Delta(0)=\Delta_0\;,~~\Delta(L)=\Delta_1\;, \label{a2}
\end{equation}
the minimum of $F^{(2)}_{\rm wire}$ is achieved when \cite{A}
\begin{equation}
\Delta(x)=\left[\Delta_0\sin\frac{L-x}{\xi}+\Delta_1\sin\frac{x}{\xi}
\exp({-i{A}_{01}})\right]\frac{\exp\,i{ a}(x)}{\sin\eta}\;,
                                                    \label{a3}
\end{equation}
where $\eta={L}/{\xi}$,
\begin{equation}
\xi\equiv\xi(T)=\sqrt{\frac{\gamma}{\alpha}}
\approx \frac{\overline{\xi}}{\sqrt{1-T/T_{c0}}} \label{a4}
\end{equation}
is the temperature dependent correlation length [here
\makebox{$\overline{\xi}\sim\xi(T=0)$]}, the function ${a}(x)$ is
defined by the integral
\begin{equation}
{ a}(x)=\frac{2\pi}{\Phi_0}\int_{0}^{x}dx'\, A_\|(x')\;, \label{a5}
\end{equation}
whereas $A_{01}$ is the value of this integral for the whole wire,
${A}_{01}={a}(L)$.

Substitution of Eq. (\ref{a3}) into the expression
for the superconducting current in the wire,
\begin{equation}
I(x)=\frac{2e}{\hbar}\gamma\,\mbox{Re}\left[\Delta^*(x)
\left(-i\frac{\partial}{\partial
x}-\frac{2\pi}{\Phi_0}A_\|\right)\Delta(x)\right]\;, \label{a7a}
\end{equation}
shows that the value of the current is constant along the wire
and is given by
\begin{equation}
I_{01}=-\frac{2e}{\hbar}\frac{\gamma}{\xi\sin\eta}\mbox{Im}
\left[\Delta_{0}\Delta_{1}^{*}e^{iA_{01}}\right]\;.  \label{a7b}
\end{equation}

On the other hand substitution of Eq. (\ref{a3}) into Eq.
(\ref{a1a}) gives a simple quadratic form of $\Delta_0$ and
$\Delta_1$ \cite{SL}:
\begin{equation}
F^{(2)}_{\rm wire}(\Delta_0,\Delta_1,{A}_{01})
=F_2\cdot\left[\cos\eta\left(|\Delta_0|^2+|\Delta_1|^2\right)
-\langle\Delta_0|\Delta_1\rangle \right]\;, \label{a6}
\end{equation}
where $F_2=\gamma/(2\xi\sin\eta)$ and
\begin{equation}
\langle\Delta_j|\Delta_k\rangle=\Delta_j\Delta_k^* e^{i{A}_{jk}}
+\Delta_j^*\Delta_k e^{-i{A}_{jk}}\;.                \label{a6b}
\end{equation}

\subsection{An arbitrary network}

The function $F^{(2)}_{\rm wire}$ defined by Eq. (\ref{a6}) can be
then used to express the harmonic part of a free energy of a
superconducting wire network $F^{(2)}_{\rm nw}$ in terms of the
values of the superconducting order parameter $\Delta_{j}$ in its
nodes $j$,
\begin{equation}
F^{(2)}_{\rm nw}=\sum_{(jk)}F^{(2)}
_{\rm wire}(\Delta_j,\Delta_k,{A}_{jk})\;.
                                                        \label{a8}
\end{equation}
Here the summation is performed over all links $(jk)$ of a network.
In the following we assume that all the links are
identical and, therefore, the function $F^{(2)}_{\rm wire}(\Delta_j,
\Delta_k,{A}_{jk})$ is the same for all the links.

In the case of a network formed by identical plaquettes it is
convenient to express the value of perpendicular external magnetic
field $H$ in terms of the number of flux quanta per single
plaquette: $f=HS/\Phi_0$ (here $S$ is the area of a plaquette).
Then the directed summation of the variables ${A}_{jk}\equiv
-{A}_{kj}$ along the perimeter of each plaquette in positive
direction (denoted below as $\sum_{\Box}$) should give
\begin{equation}
\sum_{\Box}{A}_{jk}=2\pi f\;.                  \label{a9}
\end{equation}
From the form of Eq. (\ref{a6}) it is evident that the shift of
$f$ by an integer or its reflection with respect to $f=1/2$
($f\rightarrow 1-f$) do not change the form of the expression for
free energy (or can be taken care of by a redefinition of
variables), and, therefore, it is sufficient to analyze the
interval $0\leq f\leq 1/2$. By the analogy with frustrated $XY$
models \cite{TJ} a network with the maximal irreducible value of
$f$, that is with $f=1/2$, can be called a fully frustrated
network.

When fluctuations are completely neglected, the magnetic field
dependence of the superconducting transition temperature in a
network $T_c(f)$ can be found by looking when (with the decrease
of temperature) the quadratic form defined by Eqs.
(\ref{a6})-(\ref{a8}) looses its positiveness.
To this end one has
to analyze the system of equations obtained by the variation of
$F^{(2)}_{\rm nw}$ with respect to $\Delta_{k}^*$,
\begin{equation}
\sum_{j=j(k)}\left[\Delta_k\cos\eta-\Delta_{j}
e^{i{A}_{jk}}\right]=0\;,                        \label{a10}
\end{equation}
where $j(k)$ are the nodes connected with $k$ by the links of a
network (in the following we call them the nearest neighbors of
$k$). The same equations can be derived \cite{A,dG} directly in
the framework of the continuous description without explicit
calculation of $F^{(2)}_{\rm wire}(\Delta_{j},\Delta_{k},A_{jk})$.
Multiplication of Eq. (\ref{a10}) by $\Delta_{k}^*$ with
subsequent extraction of the imaginary part allow to obtain the
current conservation equation,
\begin{equation}
\sum_{j=j(k)}I_{jk}=0\;. \label{a12}
\end{equation}

The form of Eq. (\ref{a10}) coincides \cite{A} with that of the
Schr\"{o}dinger equation for a single electron hopping between the
sites of the lattice with the same geometry in the presence of
external magnetic field. As a consequence, $T_c(f)$ can be related
with $\epsilon_0(f)$, the largest eigenvalue of the
Schr\"{o}dinger equation in the same field. For a network whose
nodes are all characterized by the same coordination number $z$
this relation can be written \cite{PCRV} as
\begin{equation}
\frac{T_{c0}-T_c(f)}{T_{c0}} =
\left[\frac{\overline{\xi}}{L}
\arccos\frac{\epsilon_0(f)}{z}\right]^2\;.
                                                     \label{a11}
\end{equation}
Starting from the work of 
Hofstadter \cite{H} (who considered the case of a square lattice),
the spectrum of the Schr\"{o}dinger equation for a single electron
hopping problem in the presence of external magnetic field has
been extensively studied for various types of two-dimensional
lattices including triangular \cite{CW}, honeycomb \cite{R}, dice
\cite{VMD} and {\em kagome} \cite{LN,XPCH} lattices.

The structure of the superconducting state in the network just below
$T_c(f)$ is determined by the structure of the eigenfunction
corresponding to $\epsilon_0(f)$ \cite{WRP}.
The conditions for the applicability of the mean field approach
are discussed in Appendix A.

\subsection{A network with a dice lattice geometry}

Dice lattice \cite{HC,S} is formed by the sites with the
coordination numbers 3 and 6 in such a way that each bond connects
two sites with different coordination numbers (see Fig. 1). Below,
when discussing a dice lattice, we denote the three-fold
coordinated sites $k$ and the six-fold coordinated sites $j$.
Thus, the bond $(jk)$ of a dice lattice connects the six-fold
coordinated site $j$ with the three-fold coordinated site $k$.

When considering the problem on a dice lattice it is convenient to
simplify the quadratic form (\ref{a8}) by minimizing it with
respect to all variables $\Delta_k$ defined on the three-fold
coordinated sites. Substitution [from Eq. (\ref{a10})] of
\begin{equation}
\Delta_k=\frac{1}{3\cos\eta}\sum_{j=j(k)} \Delta_{j} e^{i{A}_{jk}}
                                               \label{b0}
\end{equation}
into Eqs. (\ref{a6})-(\ref{a8}) then gives:
\begin{eqnarray}
F^{(2)}_{\rm nw}
&=&\frac{F_2}{3\cos\eta}\sum_{(jj')}\left[(3\cos^2\eta-1)
\left(|\Delta_j|^2+|\Delta_{j'}|^2 \right) \right. \nonumber \\
& & \left.\mbox{}-2\cos(\pi f)
\left(e^{i{A}_{jj'}}\Delta_j\Delta_{j'}^*+\mbox{c.c.}\right)
\right]                                                 \label{b1}
\end{eqnarray}
where the summation is performed over the pairs $(jj')$ of
nearest neighbors on the triangular lattice formed by the six-fold
coordinated sites, whereas variables
\begin{equation}
{A}_{jj'}=\left[({A}_{jk'}+{A}_{k'j'})+
({A}_{jk''}+{A}_{k''j'})\right]/2                  \label{b2}
\end{equation}
(where $k'$ and $k''$ are the two three-fold coordinated sites
belonging to the same rhombus as $j$ and $j'$) are the averages of
$A_{jj'}$ on the two shortest paths on a network connecting the
nodes $j$ and $j'$. It follows from Eq. (\ref{a9}) that the
variables ${A}_{jj'}\equiv -{A}_{j'j}$ have to satisfy the
constraint
\begin{equation}
\sum_{\Box}{A}_{jj'}=3\pi f\;             \label{b3}
\end{equation}
on all plaquettes of the triangular lattice. The form of Eqs.
(\ref{b1}) and (\ref{b3}) suggests that for $0\leq f\leq 1/2$ the
problem of finding $T_c(f)$ on a dice lattice is reduced to
analogous problem on a triangular lattice with $f$ multiplied by
$3/2$ and a different value of $\eta$. Accordingly, the relation
between the critical temperatures (expressed in terms of $\eta$)
in the two cases is given by
\begin{equation}
3\cos^2\eta_c^{}(f)-1
=2\cos(\pi f)\cos\eta_c^\bigtriangleup (3f/2)\;.         \label{b4}
\end{equation}
Analogous relation between the single electron spectra on dice and
triangular lattices has been derived in Ref. \onlinecite{VMD}.

Quite remarkably, for $f=1/2$ the non-diagonal coupling
in Eq. (\ref{b1}) completely disappears,
which allows immediately to conclude that
\begin{equation}
\eta_c\left({1}/{2}\right) =\arccos({1}/{\sqrt{3}})\approx
0.9553\;. \label{b5}
\end{equation}
This absence of coupling between different variables $\Delta_j$
can be understood as a manifestation of the extremely localized
nature of the highly degenerate eigenfunctions \cite{VMD}
corresponding to the largest eigenvalue of the single electron
Hamiltonian.

As a consequence, for $f=1/2$ the value of $F^{(2)}_{\rm nw}$
turns out to be exactly the same for any set of the variables
$\Delta_j$ satisfying the normalization condition
\begin{equation}
\frac{1}{N}\sum_{j}|\Delta_j|^2=\Delta^2\;,              \label{b6}
\end{equation}
where $N$ is the number of the six-fold coordinated sites in the
network with appropriately chosen periodic boundary conditions
(the total number of sites being $3N$). Accordingly, to find the
structure of the superconducting state in a fully frustrated wire
network with the dice lattice geometry (which is the main subject
of this work) one has to minimize the fourth-order contribution to
free energy,
\begin{equation}
F^{(4)}_{\rm nw}=\sum_{(jk)}F^{(4)}_{\rm wire}(\Delta_j,\Delta_k,
{A}_{jk})\;,                                       \label{c5}
\end{equation}
[where $\Delta_k$ is given by Eq. (\ref{b0})] with respect to the
whole set of the variables $\Delta_j$ satisfying the constraint
(\ref{b6}), which fixes also the value of $F^{(2)}_{\rm nw}$. For
$0<\eta-\eta_c\ll 1$
\begin{equation}
F^{(2)}_{\rm nw}\approx -12NF^{}_2\Delta^2(\sin\eta_c)
(\eta-\eta_c)\;,                                 \label{c6}
\end{equation}
where we have kept only the lowest order term of the expansion
with respect to $\eta-\eta_c$.

At the conceptual level, this task can be considered as analogous
to finding the structure of the vortex lattice which minimizes the
forth-order contribution to the free energy of a bulk
superconductor just below $H_{c2}$. In this problem (first
analyzed by Abrikosov \cite{Abr}), the harmonic contribution to
free energy is degenerate with respect to a huge number of
continuous variables, the positions of the order parameter
singularities, whereas in the present problem  a huge continuous
degeneracy of the harmonic problem is related with variables
$\Delta_j$.

\section{Fourth-order contribution to free energy}
\label{quarticFE}
\subsection{A single wire}

Substitution of Eq. (\ref{a3}) into Eq. (\ref{a1b})
describing the fourth-order contribution
to the free energy of a superconducting wire gives
\begin{eqnarray}
F^{(4)}_{\rm wire}(\Delta_{j},\Delta_{k},A_{jk}) & = & F_4 \left[
I_4\cdot\left(|\Delta_j|^4+|\Delta_k|^4\right)\right. \label{c2}\\
& + &
2I_3\cdot\left(|\Delta_j|^2+|\Delta_k|^2\right)\langle\Delta_j|\Delta_k\rangle
                          \nonumber                 \\ & + &
I_2\cdot\left.\left(2|\Delta_j|^2|\Delta_k|^2
+\langle\Delta_j|\Delta_k\rangle^2\right)\right]       \nonumber
\end{eqnarray}
where $F_4=\beta L/4$ and the numerical constants $I_n$ (with
$n=2,3,4$) are given by the integrals
\begin{equation}
I_n = \int_{0}^{1}dt\,\frac{\sin^n(\eta t)\sin^{4-n}[\eta(1-t)]}
          {\sin^4\eta}\;.                            \label{c3}
\end{equation}
When we are interested in the structure of the superconducting
phase just below $T_c(f)$, the comparison of the forth order terms
in the free energy of different states should be made by
calculating them at $T=T_c(f)$. Thus, in the following we will
need the values of $I_n$ at
$\eta=\eta_c(1/2)=\arccos(1/\sqrt{3})$, which are
\begin{equation}               \label{c4}
\begin{array}{rcl}
I_2 & = & ({15}-{9\sqrt{2}}{\eta_c}^{-1})/32
\approx 0.0524\;,  \\
I_3 & = & \sqrt{3}({7\sqrt{2}}{\eta_c}^{-1}-9)/32
\approx 0.0737\;,  \\
I_4 & = & ({27}-{13\sqrt{2}}{\eta_c}^{-1})/32
\approx 0.2424\;.
\end{array}
\end{equation}

\subsection{A tripod of three wires}

For $f=1/2$
the contribution to $F^{(4)}_{\rm nw}$ from a tripod formed by the
three links $(j_a k)$ [where $j_a\equiv j_a(k)$ with $a=1,2,3$ are
the three nearest neighbors of $k$ numbered in the positive
direction] after the substitution of Eq. (\ref{b0}) can be
rewritten as
\begin{eqnarray}
\lefteqn{F^{(4)}_{{\rm nw}}(k) = } \nonumber \\
 & &
F_4  \left[
\frac{3\nu_1+\nu_2}{4}\sum_{a=1}^{3}|\Delta_{j_a}|^{4}+
\frac{\nu_1+\nu_2}{4}\left(\sum_{a=1}^{3}|\Delta_{j_a}|^{2}\right)^{2}+
\right. \nonumber \\
&   &
+\nu_1\left(\sum_{a=1}^{3}|\Delta_{j_a}||\Delta_{j_{a+1}}|
\sin\chi_{j_{a}j_{a+1}}\right)^{2}+ \nonumber \\
&   & +\nu_2\sum_{a=1}^{3}\left(|\Delta_{j_a}||\Delta_{j_{a+1}}|
\sin\chi_{j_{a}j_{a+1}}\right)^{2} - \label{ordre4} \\
&   & -\nu_1\sum_{a=1}^{3}|\Delta_{j_a}||\Delta_{j_{a+1}}|
|\Delta_{j_{a+2}}|^{2}\sin\chi_{j_{a}j_{a+1}} - \nonumber
\end{eqnarray}
\vspace*{-5mm}
\[
-(\nu_1+\nu_3)\sum_{a=1}^{3}|\Delta_{j_a}||\Delta_{j_{a+1}}|
(|\Delta_{j_a}|^{2}+|\Delta_{j_{a+1}}|^{2})\left.
\sin\chi_{j_{a}j_{a+1}}\right]\;,
\]
where
\begin{eqnarray*}
\nu_1 & = &   \frac{4}{3}I_{2}
              +\frac{16}{3\sqrt{3}}I_{3}+\frac{4}{3}I_{4}\:, \\
\nu_2  & = &  \frac{4}{3}I_{2} \:,\\
\nu_3  & = &  \frac{8}{3}I_{2}+\frac{4}{\sqrt{3}}I_{3} \:,
\end{eqnarray*}
whereas $\chi_{jj'}$ are the gauge-invariant phase differences
between the phases $\varphi_{j}$ of the order-parameter
\makebox{$\Delta_{j}\equiv|\Delta_{j}|\exp(i\varphi_{j})$} at
neighboring six-fold coordinated sites,
\begin{equation}                               \label{d3}
\chi_{jj'}=\varphi_{j'}-\varphi_{j} -{A}_{jj'}\equiv -\chi_{j'j}\;.
\end{equation}
It follows from Eq. (\ref{b3}) that for all tripods, or, in other
words, for all plaquettes of the triangular lattice formed by the
six-fold coordinated sites,
\begin{equation}                               \label{d3a}
\sum_{a=1}^{3}\chi_{j_a j_{a+1}} =-\sum_{a=1}^3 A_{j_a
j_{a+1}}=-3\pi/2\;.
\end{equation}
Since each plaquette of this lattice has a particular three-fold
coordinated site in its center, the index $k$ numbering such sites
can be also used for numbering triangular plaquettes.

In the cyclic sums in Eq. (\ref{ordre4}) and analogous sums below
$j_4 \equiv j_1$. The last term in Eq. (\ref{ordre4}) can be
omitted, since during summation over the whole lattice the two
tripods adjacent to any link $(j_a j_b)$ always yield opposite
contributions.

\subsection{The equal amplitude hypothesis}

Let us now introduce the additional assumption (whose
self-consistency is established in Appendix~\ref{Consistency}) that
the absolute values of the variables $\Delta_j$ are the same for all
six-fold coordinated sites $j$,
\begin{equation}
\Delta_j=\Delta\exp(i\varphi_j)\;,                  \label{d1}
\end{equation}
where $\Delta$ is real.
In that case the contributions to
\begin{equation}
F^{(4)}_{\rm nw}=\sum_{k}F^{(4)}_{\rm nw}(k)       \label{d3f}
\end{equation}
from the next but last term in Eq. (\ref{ordre4})
coming from the neighboring tripods also cancel each other,
and the expression for $F^{(4)}_{\rm nw}$ is reduced to
\begin{equation}
F^{(4)}_{\rm nw}=F_4\Delta^4\sum_{k}\left[\nu_0 + V(\{\chi_{j_a
j_{a+1}}\})\right]\;,                           \label{d3b}
\end{equation}
where $\nu_0=(9/2)\nu_1+3\nu_2$,
\begin{equation}                                \label{d4}
V(\{\chi\})=
V(\chi_1,\chi_2,\chi_3)\equiv \nu_1\left(\sum_{a=1}^{3}\sin\chi_a
\right)^2 +\nu_2\sum_{a=1}^{3}\sin^2\chi_a \;
\end{equation}
and
\begin{equation}                                \label{d4b}
\sum_{a=1}^{3}\chi_{j_a j_{a+1}}=\pi/2 ~\mbox{(mod $2\pi$)}\;.
\end{equation}
Since $F^{(4)}_{\rm nw}$ is invariant with respect to the shift of
any of the variables $\chi_{jj'}$ by a multiple of $2\pi$, here
and below we for convenience assume that they all are reduced to
the interval $[-\pi,\pi]$, in accordance with which the right hand
side of Eq. (\ref{d4b}) is written as $\pi/2$ (mod $2\pi$) instead
of $-3\pi/2$, as it would follow from Eq. (\ref{d3a}).

At the temperature of the superconducting transition in a fully
frustrated wire network [that is, at $\eta=\eta_c=\eta_c(1/2)$]
the values of the coefficients $\nu_n$ are given by
\begin{equation}                                    \label{d5}
\begin{array}{rcl}
\nu_1 & = & \frac{1}{4}(1+\sqrt{2}\eta_c^{-1})\approx  0.6201\;,\\
\nu_2 & = & \frac{1}{8}(5-3\sqrt{2}\eta_c^{-1}) \approx 0.0699\;, \\
\nu_0 & = & 3 \;.
\end{array}
\end{equation}

\section{Minimization of the fourth-order contribution to free energy}
\label{Minimization}

\subsection{A single triangle}

It is well known that the ground state \cite{DCJW,MS} of the
antiferromagnetic $XY$ model with triangular lattice can be found
by minimizing the energy separately for each triangular plaquette
and then matching these solutions with each other.

For $\nu_1,\nu_2>0$ the minimum of $V(\chi_1,\chi_2,\chi_3)$  on
an isolated triangle [under the constraint of the form
(\ref{d4b})] is achieved when two of the arguments of
$V(\chi_1,\chi_2,\chi_3)$ are equal to each other, for example
\begin{equation}
\chi_1=\chi_2=-\psi(t)\;,~\chi_3=\pi/2+2\psi(t)\;, \label{d6a}
\end{equation}
another solution with the same value of $V(\chi_1,\chi_2,\chi_3)$
being
\begin{equation}
\chi_1=\chi_2=-[\pi-\psi(t)]\;,~ \chi_3=\pi/2-2\psi(t)\;,
                                                  \label{d6b}
\end{equation}
where $t=\nu_2/\nu_1$ and
\begin{equation}
 \psi(t)=\arcsin{\frac{\sqrt{12+4t+t^2}-2+t}{4(1+t)}}\;.
\end{equation}
With increase of $t$ from zero to infinity $\psi(t)$ continuously
increases from $\arcsin{[(\sqrt{3}-1)/2]}\approx \pi/8$ to
$\pi/6$.

In an infinite system each variable $\chi_{jj'}$ belongs
simultaneously to two triangles, but enters the function
$V(\chi_1,\chi_2,\chi_3)$ on these two triangles with the opposite
signs. Comparison of Eqs. (\ref{d6a}) and (\ref{d6b}) with each
other allows to conclude that it is impossible to minimize
$F^{(4)}_{\rm nw}$ by minimizing $V(\chi_1,\chi_2,\chi_3)$
separately for each triangle.

\subsection{A periodic solution}

When variables $\chi_{jj'}$ are reduced to a finite interval, the
average value of $\sum_{a=1}^{3}\chi_{j_a j_{a+1}}$ should be
equal to zero. This can be achieved if on one quarter of triangles
the right hand side of Eq. (\ref{d4b}) is equal to $-3\pi/2$,
whereas on all remaining triangles it is really equal to $\pi/2$.
Accordingly, the minimal supercell of a periodic set of variables
$\chi_{jj'}$ should consist of four triangles.

The four-triangle supercell with the most symmetric (triangular)
shape, but with the most general structure allowing for
construction of a periodic state by a periodic repetition of this
supercell, is shown in Fig. 2a. It can be described by the six
variables $\chi_i$ (with $i=1,\ldots,6$) defined as
shown in Fig. 2a and satisfying three independent constraints of
the form (\ref{d4b}), which can be chosen to be
\begin{eqnarray}
\chi_{1}+\chi_{2}+\chi_{3} & = & {\pi}/{2}\;, \label{e2a}\\
-\chi_{3}+\chi_{4}+\chi_{5} & = & {\pi}/{2}\;, \label{e2b} \\
-\chi_{1}-\chi_{5}+\chi_{6} & = & {\pi}/{2}\;, \label{e2c}
\end{eqnarray}
the fourth constraint,
\begin{equation}
-\chi_{2}-\chi_{4}-\chi_{6}   =  -{3\pi}/{2}\;,\label{e2d}
\end{equation}
following automatically from Eqs. (\ref{e2a})-(\ref{e2c}).

\begin{figure}[bt]
\includegraphics[width=45mm]{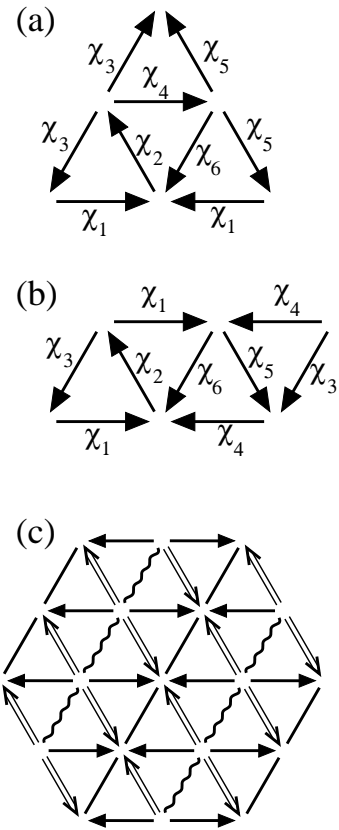}
\caption[Fig. 2]
{Construction of  periodic patterns
minimizing the fourth-order contribution to free energy.
(a) A possible choice for the most symmetric four triangle supercell.
(b) An alternative four triangle supercell.
(c) The structure of the periodic solution
obtained from free energy minimization
with the supercell shown in (a) or (b). Simple arrows correspond to
phase differences $\chi_{jj'}$ equal to $\pm \pi/4$, double arrows
to $\pm 3\pi/4$, simple lines to $0$ and wiggly lines to $\pi$.}
\label{fig2}
\end{figure}

The minimization of $F^{(4)}_{\rm nw}$ for this supercell with
respect to the remaining three degrees of freedom shows that for
$\chi_{i}\in[-\pi,\pi]$ the minimum is achieved when
\begin{equation}
\begin{array}{ll}
\chi_{1}=\chi_{4}=-\pi/4   \;,~~ & \chi_{3}=0 \;,  \\
\chi_{2}=\chi_{5}=3\pi/4 \;,   & \chi_{6} =\pi\;,
\end{array}                                           \label{e3}
\end{equation}
or in one of the five other states which can be constructed from
this state by permutations of the variables $\chi_{i}$. In
all these states on all triangles
\begin{equation}
\sum_{a=1}^3\sin{\chi_{j_a j_{a+1}}}=0\;,         \label{e3b}
\end{equation}
which means that on each triangle the first term of $V(\{\chi\})$
reaches its absolute minimum ({\em i.e.}, is equal to zero).
Accordingly, the value of $F^{(4)}_{\rm nw}$ in these states does
not depend on $\nu_1$,
\begin{equation}
F^{(4)}_{\rm nw}=2(\nu_0+{\nu_2})F_4\Delta^4 N\;.
                                                  \label{e4}
\end{equation}

Note that the supercell defined by Eqs. (\ref{e3}) consists  of
two pairs of equivalent (if one takes into account the equivalence
of $\pi$ and $-\pi$) triangles. Thus the actual size of the
supercell has turned out to be two times smaller than it has been
initially conjectured. But there was no way to predict this
without really performing the minimization for the four-triangle
supercell.

The same solution (whose structure is shown in Fig 2c) can be also
found by starting from the assumption that a periodic state is
formed with the help of the four-triangle supercell with the
different shape shown in Fig. 2b. In that case the constraints
(\ref{e2a}) and (\ref{e2b}) retain their form, whereas in Eqs.
(\ref{e2c}) and (\ref{e2d}) one should interchange $\chi_1$ and
$\chi_4$. For this supercell the minimum of $F^{(4)}_{\rm nw}$
(for not too large ratio $\nu_2/\nu_1$) is again achieved in the
solution described by Eqs. (\ref{e3}) (or other equivalent
solutions).

It follows from Eq. (\ref{b0}) that for $|\Delta_{j_a}|=\Delta$
\begin{equation}
|\Delta_{k}|^2=\frac{\Delta^2}{3}\left(3-2\sum_{a=1}^{3}
\sin\chi_{j_a j_{a+1}}\right)\;.            \label{e5}
\end{equation}
Substitution of Eq. (\ref{e3b}) into Eq. (\ref{e5})  allows
immediately to conclude that in the solution which we have found
the absolute value of the order parameter on all three-fold
coordinated sites has the same value as on the six-fold
coordinated sites,
\begin{equation}
|\Delta_{k}|=|\Delta_{j}|=\Delta\;.          \label{e9}
\end{equation}

In addition to gauge-invariant variables $\chi_{jj'}$ defined on
the bonds of triangular lattice, one can, naturally, also introduce
the gauge-invariant phase differences defined on the bonds
of the original dice lattice,
\begin{equation}
\theta_{jk}=\varphi_k-\varphi_j-A_{j k}\;,    \label{thetajk}
\end{equation}
where $\varphi_k\equiv\arg{(\Delta_k})$.
As a consequence of Eq. (\ref{a9}), the variables
$\theta_{jk}=-\theta_{kj}$, which we assume to be reduced to the
interval $(-\pi,\pi)$, have to satisfy the constraint
\begin{equation}
\sum_{\Box}{\theta}_{jk}=\pm\pi\;. \label{sumth}
\end{equation}
When Eq.  (\ref{e9}) is fulfilled, the expression (\ref{a7b}) for
the current in a link is reduced to
\begin{equation}
I_{jk}=I_0\sin\theta_{jk}\;,                    \label{f4}
\end{equation}
where
\begin{equation}
I_0=\frac{2e}{\hbar}\frac{\gamma\Delta^2}{\xi\sin\eta}\;.
                                                \label{f5}
\end{equation}

Calculation of
\begin{equation}
\Delta_{j}\Delta_{k}^*\exp({i{A}_{j k}})
=\Delta^2\exp(-i\theta_{jk})                   \label{e6}
\end{equation}
with the help of Eq. (\ref{b0}) demonstrates that in the
considered state the variables $\theta_{jk}$ have the same three
values (up to a permutation and a simultaneous change of sign),
\begin{equation}
\theta_{j k}= \pm \theta_{1}\;,~ \pm \theta_{2}\;,~ \mp
\theta_{3}\;,                               \label{e7}
\end{equation}
on all tripods forming dice lattice. These values have to satisfy
the constraints,
\begin{equation}
\theta_2-\theta_1=\pi/4\;,~~ \theta_1+\theta_3=\pi/2\;,~~
\theta_2+\theta_3=3\pi/4\;,                           \label{e8b}
\end{equation}
leading to the automatic fulfillment of Eqs. (\ref{sumth}), as
well as the current conservation equation,
\begin{equation}                        \label{currcons}
  \sin\theta_1+\sin\theta_2=\sin\theta_3\;,
\end{equation}
which follows from Eq. (\ref{f4}). As a consequence, they turn out
to be exactly the same,
\begin{equation}
\theta_{1,3}=\arccos{\left({1}/{\sqrt{3}}\mp{1}/{\sqrt{6}}\right)}\;,~
\theta_{2}=\arccos\left({{1}/{\sqrt{3}}}\right)\;,
                                                      \label{e8}
\end{equation}
as in the ground state of the fully frustrated $XY$ model (FFXYM)
with a dice lattice \cite{K01}, for which the current conservation
equation also has the form (\ref{currcons}).

Thus, in terms of $\theta_{jk}$, the state which we have found (it
is schematically shown in Fig. 3a) has exactly the same structure
as one of the ground states of the FFXYM with a dice lattice. It
has to be emphasized that the reasons for that are more subtle
than a simple reduction of one model to the other. Firstly, in the
case of a superconducting wire network the relation (\ref{e9}),
the form of which seems to imply a possible reduction to $XY$
model, is valid only in the minimum of free energy. Secondly, the
substitution of Eq. (\ref{e9}) into Eq. (\ref{c2}) gives
\begin{equation}
F^{(4)}_{\rm wire}(\theta_{jk})
=F_1[2(I_4+I_2)+8I_3\cos\theta_{jk}+4I_2\cos^2\theta_{jk}]\;,
                                                  \label{e10}
\end{equation}
whereas in the case of the FFXYM in the expression for the energy
of a bond the term proportional to $\cos^2{\theta}$ is simply
absent, whereas the term proportional to $\cos{\theta}$ has a
coefficient of the opposite (negative) sign. Thus the two models
do not become equivalent even if Eq. (\ref{e9}) is artificially
introduced as an additional assumption.

\begin{figure*}[bt]
\includegraphics[width=130mm]{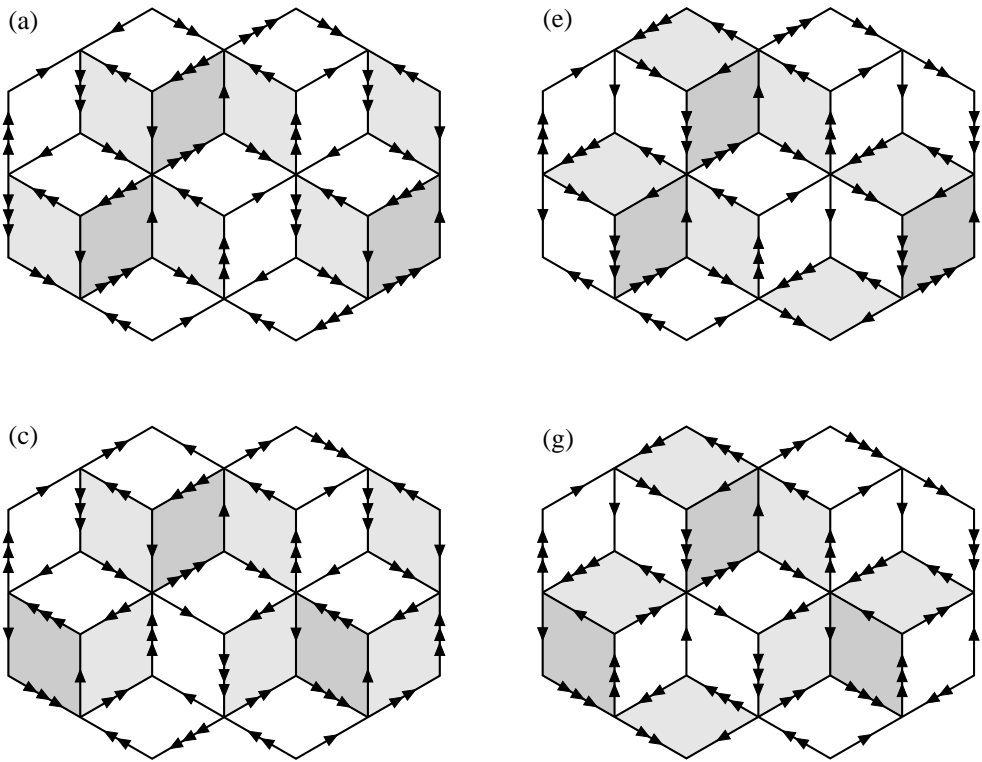}
\caption{\label{fig3}
The four periodic states generating the class of
degenerate states discussed in the text by adjunction
of domain walls. The plaquettes with positive vorticities
are shaded, and the three types of arrows correspond to the three
possible gauge-invariant phase differences $\theta_{1}$, $\theta_{2}$,
$\theta_{3}$ (modulo $2\pi$). Labels are chosen to match those
introduced in Ref.~\onlinecite{K01}.}
\end{figure*}

\subsection{Additional degeneracy}

The ground state of the FFXYM with a dice lattice is known to
possess a well-developed accidental degeneracy, which can be
described in terms of the formation of a network of zero-energy
domain walls \cite{K01} on the background of a periodic state.
This construction can start from any of the four periodic states
shown in Fig. 3 and allows to obtain, in particular, the three
other states shown in that figure by introduction of such domain
walls.

For example, both state (c) and state (e) can be obtained from
state (a) by introduction of a dense sequence of parallel
zero-energy domain walls (of two different types). In that
language state (g) can be described as the dense network of
intersecting domain walls of two types. On the other hand, if one
starts the construction from state (c), both state (a) and state
(g) are formed by introduction of parallel domain walls, whereas
the network of two types of walls corresponds to state (e).

The same set of states (described in a more detail in Ref.
\onlinecite{K01}) minimizes $F^{(4)}_{\rm nw}$ for given $\Delta$.
In Appendix B we check that all these states are extremal not only
when one assumes $|\Delta_{j}|=\mbox{const}$, but also in the absence
of this constraint.
The accidental degeneracy related to formation of zero-energy
domains walls gives rise to residual entropy proportional to the
linear size of the system.

In the framework of the description of different states in a
network in terms of the gauge-invariant phase differences
$\theta_{jk}\in(-\pi,\pi)$,
all rhombic plaquettes $\alpha$ of a dice lattice
can be considered as occupied by positive and negative
half-vortices, whose vorticities $m_\alpha=\pm 1/2$ are given by
\begin{equation}
m_\alpha=\frac{1}{2\pi}\sum_{\Box}\theta_{jk}=\pm\frac{1}{2}\;.
\label{m_A}
\end{equation}
In the family of states minimizing $F^{(4)}_{\rm nw}$ the
half-vortices of the same sign always form triads with one
"central" and two "edge" vortices \cite{K01}. The formation of
domain walls which cost no free energy is related to the changes
in the orientation and/or in the shape of these triads, but does
not lead to formation of vortex clusters of any other size.

At low temperatures ($T\ll T_{c0}$) the free energy of a fully
frustrated wire network with the dice lattice geometry (which then
can be described in terms of the London approximation) is minimal
for the same set of states, but with the slightly different values
\cite{K01} of $\theta_{a}$
\begin{equation}
\theta_{1}={\pi}/{12}\,,~~\theta_{2}={\pi}/{3}\,,~~
\theta_{3}={5\pi}/{12}\;,                        \label{e11}
\end{equation}
satisfying, nonetheless, the same constraints (\ref{e8b}).

\subsection{Alternative solution}

The analysis of the supercell shown in Fig. 2b allows also to find
a state which minimizes $F^{(4)}_{\rm nw}$ for large $\nu_2$. In
the notation of Fig. 2b the structure of this state is given by
\begin{equation}
\chi_1=\chi_2=\chi_3=-\chi_4=\pi/6; ~~\chi_5=-\chi_6=5\pi/6\;.
                                                  \label{12}
\end{equation}
It minimizes the value of the second term in Eq. (\ref{d4})
separately for each triangle, the value of $F^{(4)}_{\rm nw}$
being
\begin{equation}
F^{(4)}_{\rm nw}
=2\left[\nu_0+\frac{3}{4}(\nu_1+\nu_2)\right]F_4\Delta^4 N\;.
                                                  \label{e13}
\end{equation}

This alternative state is characterized by even more developed
accidental degeneracy leading to extensive residual entropy.
Namely, the value of $F^{(4)}_{\rm nw}$  does not change if at an
arbitrary number of sites $j$ the variables $\varphi_j$ are
shifted by $\pi$. Note that this property holds not only at
$\nu_1=0$, when it trivially follows from $V(\{\chi\})$ being
dependent only on $\cos{2\chi_a}$, but also at finite values of
$\nu_1$.

Comparison of Eq. (\ref{e13}) with Eq. (\ref{e4}) shows that the
values of $F^{(4)}_{\rm nw}$ in the two states become equal at
$\nu_2/\nu_1=3$, whereas in our case, according to Eqs.
(\ref{d5}), \makebox{$\nu_2/\nu_1\approx 0.1127\ll 3$}. At so low
values of the ratio $\nu_2/\nu_1$, the alternative solution
discussed in this subsection is simply unstable.

\section{Magnetic energy}
\label{Selection}
\subsection{An arbitrary network}

When currents in a two-dimensional wire network satisfy the
current conservation equations, the current in each link can be
expressed as a difference of so called mesh currents \cite{DK},
$I^m_\alpha$, associated with the plaquettes (meshes) of a network.
Namely, the current in the link $(jk)$ can be written as the
difference of the mesh currents associated with the two plaquettes
($\alpha$ and $\alpha'$) adjacent to this link,
\begin{equation}
I_{jk}=I^m_\alpha-I^m_{\alpha'}\;.              \label{f1}
\end{equation}
The magnetic energy of the currents in the network, $E_{\rm magn}$,
can be then expressed in terms of $I^m_\alpha$,
\begin{equation}
E_{\rm
magn}=\frac{1}{2c^2}\sum_{\alpha,\beta}L_{\alpha\beta}I^m_\alpha
I^m_\beta\;,                                    \label{f2}
\end{equation}
where a symmetric matrix $L_{\alpha\beta}$ is usually called the
mutual inductance matrix \cite{DK}.

The diagonal elements of this matrix describe the self-inductances
of current loops which can be associated with different plaquettes
of the network and, accordingly, have to be positive. On the other
hand, its non-diagonal elements describe the mutual inductances of
non-intersecting coplanar current loops. Magnetic fields of such loops
substract from each other and, therefore, the non-diagonal
elements of $L_{\alpha\beta}$ have to be negative.
In an infinite network the constraint
\begin{equation}
\sum_{\alpha}L_{\alpha\beta}=0\;               \label{f2b}
\end{equation}
has to be satisfied. This ensures the invariance of $E_{\rm magn}$
with respect to a possible redefinition of mesh currents,
\makebox{$I^m_\alpha\rightarrow I^m_\alpha+\delta I^m$} that
leaves the physical currents in the links, $I_{jk}$, intact. For
practical purposes it is convenient to define $I^m_{\alpha}$ in
such a way that
\begin{equation}
\sum_{\alpha}I^m_\alpha=0 \;.                   \label{f2c}
\end{equation}

The value of $L_{\alpha\beta}$ depends only on the relative
disposition of the two plaquettes $\alpha$ and $\beta$,
and in the limit of infinitely thin wires
can be found by calculating the double integral over their
perimeters $\Gamma_{\alpha}$ and $\Gamma_{\beta}$,
\begin{equation}
L_{\alpha\beta}=\oint_{\Gamma_\alpha} d{\bf
r}_\alpha\oint_{\Gamma_\beta} d{\bf r}_\beta \,\frac{1}{|{\bf
r}_\alpha-{\bf r}_\beta|}\;. \label{f3}
\end{equation}
The expression for $L_{\alpha\beta}$ given by Eq. (\ref{f3}) in
the case of $\alpha=\beta$ is logarithmically divergent, which
means that $L_{\alpha\alpha}$ has always to be calculated more
accurately, taking into account the finite width of the wires. The
same is true for the value of $L_{\alpha\beta}$ for neighboring
plaquettes (having a common link). In the case of more distant
neighbors (having only a common node or simply not touching each
other) one can use Eq. (\ref{f3}) based on the assumption of
infinitely thin wires without encountering any divergences.

Eq. (\ref{f3}) can be also rewritten as the double integral over the
areas of the plaquettes $\alpha$ and $\beta$. For $\alpha\neq \beta$
\begin{equation}
L_{\alpha\beta}=-\int_{S_\alpha} d^2{\bf
r}_\alpha\int_{S_\beta}d^2{\bf r}_\beta \,\frac{1}{|{\bf
r}_\alpha-{\bf r}_\beta|^3}\;, \label{f3b}
\end{equation}
which shows that $|L_{\alpha\beta}|$ rapidly decays with the
growth of $R_{\alpha\beta}$, the distance  between the centers of
the plaquettes $\alpha$ and $\beta$. For $R_{\alpha\beta}\gg L$
\begin{equation}
L_{\alpha\beta}\approx
-\frac{S_\alpha S_\beta}{R^3_{\alpha\beta}}\;.          \label{f3c}
\end{equation}

In any periodic state minimizing the free energy of a frustrated
network, the plaquettes with negative and positive values of
$I^m_\alpha$ regularly alternate with each other, so $E_{\rm
magn}$ (normalized, for example, per a single plaquette) is given
by a rapidly decaying sign alternating lattice sum. It allows one to
expect that the main contribution to this sum comes from its
largest terms, corresponding to the self-inductances of the
plaquettes and the mutual inductances of rather close neighbors.
Analogously, when comparing the magnetic energies of different
degenerate states minimizing the fourth order term in the free
energy, the main contribution to the difference between them can
be expected to come from the closest neighbors whose contributions
do not cancel each other identically.

Besides the proper energy of the magnetic field induced by
currents and given by Eq. (\ref{f2}),  one also has to take into
account the decrease of the superconducting free energy related to
the vector potential of this field. In the weak screening regime
the sum of these two contributions, $F_{\rm magn}$, differs from
$E_{\rm magn}$ only by sign,
\[
F_{\rm magn}=-E_{\rm magn}\,,
\]
and, therefore, one has to maximize $E_{\rm magn}$.

\subsection{Mutual inductances of dice lattice plaquettes}

\begin{figure}[b]
\includegraphics[width=52mm]{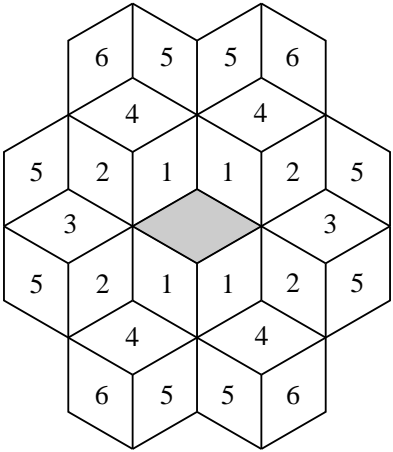}
\caption{\label{fig4}
Classification of plaquettes according to their distance
from the shaded one.}
\end{figure}

Fig. 4 introduces the classification of neighbors for rhombic
plaquettes of a dice lattice, which can be used for the natural
reordering of summation in Eq. (\ref{f2}). A chosen plaquette
(which is shaded) has four nearest neighbors (denoted by 1), four
next-to-nearest neighbors (denoted by 2), two third neighbors
(denoted by 3), {\em etc.}, as shown in Fig. 4 up to sixth
neighbors. In the following we denote the self-inductance of a
plaquette $L_0$ and the mutual inductance of a plaquette and its
$n$-th neighbor (which is a negative quantity), $L_n=-\lambda_n
L$.

For $n$ from $2$ to $5$ the calculation of $L_n$ with the help of
Eq. (\ref{f3}) or Eq. (\ref{f3b}) gives
\begin{eqnarray}
\lambda_2 & = & 4\sqrt{3}-2\sqrt{7}-2-15\ln 3
          -4\ln(1+\sqrt{3})          \\
 & & \mbox{}+\ln[8(4+\sqrt{7})^5(-1+2\sqrt{7})(1+25\sqrt{7})^2] \;, \nonumber\\
\lambda_3 & = & 8\sqrt{7}-12\sqrt{3} \\
& & \mbox{}+\ln[(27/4)(1+\sqrt{3})^{12}/(5+2\sqrt{7})^6]  \;, \nonumber \\
\lambda_4 & = & 12 -10\sqrt{3}+2\sqrt{7}-(33/2)\ln 2-(97/4)\ln 3 \\
    &  & \mbox{}+(1/2)\ln[(1+\sqrt{3})^{50}
         (4-\sqrt{7})^{31}(2+\sqrt{7})^{11}]     \;, \nonumber\\
\lambda_5 & = & 3+5\sqrt{3}-3\sqrt{7}-\sqrt{13}-12\ln 3  \\
    &  & \mbox{} +\ln[(2+\sqrt{7}) (1+2\sqrt{7})
          (5+2\sqrt{7})^3/(1+\sqrt{3})^8]  \nonumber\\
    & & \mbox{}  +(1/2)\ln[(1+\sqrt{13})(-2+\sqrt{13})^4(7+\sqrt{13})^8/2]
                    \;,                         \nonumber
\end{eqnarray}
whereas numerically
\begin{equation}                      \label{lambda}
\begin{array}{lcl}
\lambda_2 & \approx & 0.5569  \;,     \label{u2}  \\
\lambda_3 & \approx & 0.3637  \;,     \label{u3}  \\
\lambda_4 & \approx & 0.1671  \;,     \label{u4}  \\
\lambda_5 & \approx & 0.0723  \;.     \label{u5}
\end{array}
\end{equation}
Thus the decrease of $|L_n|$ with $n$ is rather fast
even for $n \sim 1$.

\subsection{The form and the magnitude of magnetic energy}

In a general situation the values of mesh currents $I^m_\alpha$
should be found from Eqs. (\ref{f1}) and, accordingly, are given by
non-local linear combinations of link currents $I_{jk}$.
Quite remarkably, substitution of Eq. (\ref{b0}) into Eq.
(\ref{a7b}) allows to find that the values
of mesh currents in a fully frustrated superconducting wire
network with dice lattice geometry in the vicinity of $T_c$
are given by the local expression,
\begin{equation}                                \label{Imjj}
  I^m_{jj'}=-\frac{2e}{\hbar}\frac{\gamma}{\xi\sin\eta}
  |\Delta_{j}||\Delta_{j'}|\frac{\cos\chi_{jj'}}{\sqrt{3}}\;,
\end{equation}
where $I^m_{jj'}$ is the mesh current in the plaquette
\makebox{$\alpha_{jj'}=(jk'j'k'')$}.

In the family of states minimizing
$F^{(4)}_{\rm nw}$ and described in Sec. IV the absolute value of
$I^m_{jj'}$ on all plaquettes acquires only two values,
\begin{equation}                              \label{IcIe}
  I^m_c=I_0/\sqrt{3}\;,~~I^m_e=I_0/\sqrt{6},
\end{equation}
where $I_0$ is given by Eq. (\ref{f5}). The choice between $I^m_c$
and $I^m_e$ is determined by whether the plaquette $\alpha_{jj'}$
is occupied by the central or edge vortex of a triad to which it
belongs, whereas the sign of $I^m_{jj'}$ by the sign of this
vortex. The ratio $g=I^m_e/I^m_c$ following from Eqs. (\ref{IcIe})
is equal to $g_1=1/\sqrt{2}$.

Substitution of Eq. (\ref{Imjj}) into Eq. (\ref{f2}) shows that
$E_{\rm magn}$ is of the fourth order in $\Delta_j$, and thus
should be added (with the negative sign) to the forth-order term,
$F^{(4)}_{\rm nw }$, which has been minimized in Sec. IV. It
follows from Eq. (\ref{Imjj}) that for $|\Delta_j|=\Delta$ the
main contribution to $E_{\rm magn}$ [which is related to
self-inductances of lattice plaquettes, $L_0\sim 8 L\ln(L/d)$,
where $d$ is the thickness of the wires], can be written as
\begin{equation}                                   \label{E0}
E^{(0)}_{\rm magn}=E_{\rm m} \sum_{(jj')}[1-\sin^2\chi_{jj'}]\;,
\end{equation}
where $E_{\rm m}={L_0 I_0^2}/{6c^2}$.

The value of the coefficient $E_{\rm m}$, which at $T=T_c$ is
given by
\begin{equation}                             \label{E_0}
  E_{\rm m}=\frac{L_0}{4}\left(\frac{2e}{\hbar c}\right)^2
  \left(\frac{\gamma}{\xi}\right)^2\Delta^4\;,
\end{equation}
should be compared with $F_4\Delta^4=({L}/{4})\beta\Delta^4$.
With the help of Eq. (\ref{Ap7}) and Eq. (\ref{Ap6}) one obtains
\begin{equation}                             \label{EmF4}
  \frac{E_{\rm m}}{F_4\Delta^4}\sim
  \frac{\sigma}{\kappa^2\xi^2(T_c)}\;,
\end{equation}
where $\kappa$ is the Ginzburg-Landau parameter of the material from
which the wires are made, and $\sigma$ their cross-section area.
Since the mean field phase transition in a fully frustrated
network takes place when $\xi(T_c)\approx L$, Eq. (\ref{EmF4}) can
be rewritten as
\begin{equation}                             \label{EmF5}
  \frac{E_{\rm m}}{F_4\Delta^4}\sim
  \frac{\sigma}{\kappa^2 L^2}\;.
\end{equation}

Thus, in order to get $E_{\rm magn}\ll F^{(4)}_{\rm nw}$ (the weak
screening regime) one should have
\[
\sigma\ll\kappa^2L^2\;.
\]
In other terms the same condition can be rewritten as
\[\Lambda_{\rm eff}[\xi(T_c)\sim L]\gg L\;,\]
where $\Lambda_{\rm eff}(\xi)\sim (\kappa\xi)^2 L/\sigma$ is the
effective penetration depth for the magnetic field in a network.
In such a case $E_{\rm magn}$ can be treated as a small
correction, which is relevant only for the removal of the
accidental degeneracy between different states minimizing
$F^{(4)}_{\rm nw}$. In the opposite limit finding the structure of
the ordered state becomes an even more complicated problem because
one has to minimize $F^{(2)}_{\rm nw}+F^{(4)}_{\rm nw}+E^{}_{\rm
magn}$ taking into account the dependence of all three terms on
the magnetic fields induced by the currents. Therefore, below we
always assume $\sigma\ll\kappa^2L^2$, that is that the wires are
thin enough.

Note, however, that the applicability of the mean field approach
requires the wires to be not too thin,
\makebox{$\sigma\gg\kappa^2L^3/\Lambda_{\rm univ}(T)$,} see
Appendix A. The two conditions on $\sigma$ are compatible provided
$L\ll\Lambda_{{\rm univ}}(T)$, which is readily satisfied in
experiments, since for a temperature around one Kelvin
$\Lambda_{{\rm univ}}(T)$ is of the order of one centimeter, see
Eq.~(\ref{Ap9}).

It is clear from the form of Eq. (\ref{E0}) that the substraction
of $E^{(0)}_{\rm magn}$ from $F^{(4)}_{\rm nw }$ does not change
the functional form of the fourth order term (expressed in the
terms of the variables $\chi_{jj'}$), but leads only to a small
increase of the coefficient $\nu_2$ in Eq. (\ref{d4}). The next
contribution to $E^{}_{\rm magn}$, which is related to mutual
inductances of neighboring plaquettes, has a more complicated
structure,
\begin{equation}                                  \label{E1}
  E^{(1)}_{\rm magn}=-\frac{L_1 I_0^2}{3}\sum_{k}\sum_{a=1}^{3}
  \cos\chi_{j_a j_{a+1}}\cos\chi_{j_{a+1}j_{a+2}}\;.
\end{equation}
However, in all the states minimizing $F^{(4)}_{\rm nw }$, the
three variables $\chi_{jj'}$ on each triangle are always given by
\begin{equation}                      \label{chi123}
\chi_{1,2,3}=\pi/2\mp\pi/2,\;\mp\pi/4,\;\pm 3\pi/4\;,
\end{equation}
as a consequence of which the sum over $a$ in Eq. (\ref{E1}) is
equal to $-1/2$ for all $k$ and, therefore, the substraction of
$E^{(1)}_{\rm magn}$ from $F^{(4)}_{\rm nw }-E^{(0)}_{\rm magn}$
also does not lift the degeneracy. To be sure of that we have
checked also that all the states minimizing $F^{(4)}_{\rm nw }$
remain extremal with respect to variations of $\varphi_j$ even
when one includes into analysis the extra terms obtained by the
variation of $E^{(1)}_{\rm magn}$.

\subsection{Selection of the state by magnetic energy}

In a more general situation, when the ratio $g=I^m_e/I^m_c$ is
kept as a free parameter, the first two contributions to the
magnetic energy can be written as
\begin{eqnarray}                              \label{Em12}
E^{(0)}_{\rm magn}
& = &\frac{L_0}{2}\left(\frac{I^m_c}{c}\right)^2 (1+2g^2)N \;,\\
E^{(1)}_{\rm magn}
& = &-L_1\left(\frac{I^m_c}{c}\right)^2 2g^2N \;,
\end{eqnarray}
and, naturally, are also the same for all the states minimizing
$F^{(4)}_{\rm nw}$.

The difference appears when considering $E^{(2)}_{\rm magn}$, the
contribution to $E_{\rm magn}$ coming from the mutual inductances
of the plaquettes which are next-to-nearest neighbors of each
other. Comparison of $E^{(2)}_{\rm magn}$ for the four periodic
states the structure of which is shown in Fig. 3 shows that for
any $g$ with $|g|<1$ the minimum of $E^{(2)}_{\rm magn}$ is
achieved in the state (e) and the maximum in the state (c).

The non-equivalent contributions to magnetic energy
of different periodic states can be characterized
by the dimensionless parameter $\epsilon$ defined by the relation
\begin{equation}                                \label{eps}
E_{\rm magn}=E^{(0)}_{\rm magn}+E^{(1)}_{\rm magn}+
\epsilon_{\rm}\frac{L(I^m_c)^2}{c^2}N\;.
\end{equation}
The values of $\epsilon$ for basic periodic states can be then
expressed in terms of the coefficients $\lambda_n$ as
\begin{eqnarray}
\epsilon_a & = & 2g^2\lambda_2-(1+2g^2)\lambda_3+(2-4g^2)\lambda_4 \nonumber\\
           &   & \mbox{}+4g^2\lambda_5+\ldots\;,           \label{epsa}\\
\epsilon_c & = & (1+g^2)\lambda_2-g^2\lambda_3+(1-3g^2)\lambda_4 \nonumber\\
           &   & \mbox{}-(2-2g^2)\lambda_5+\ldots\;,       \\
\epsilon_e & = & -2g^2\lambda_2-(1-2g^2)\lambda_3+2\lambda_4
                 -4g^2\lambda_5+\ldots\;,           \\
\epsilon_g & = & (1-3g^2)\lambda_2+g^2\lambda_3+(1+g^2)\lambda_4 \nonumber \\
           &   & \mbox{}-(2-2g^2)\lambda_5+\ldots\;.       \label{epsg}
\end{eqnarray}
Substitution of the values of $\lambda_n$ given by Eqs.
(\ref{lambda}) into Eqs. (\ref{epsa})-(\ref{epsg}) shows that for
$|g|<1$ the state (c) maximizes the $n$-th approximation to
$\epsilon$,
\begin{equation}
\epsilon^{(n)}_{\rm}
=\frac{c^2}{L(I^m_c)^2N}\sum_{m=2}^{n}E^{(m)}_{\rm magn}
\end{equation}
not only for $n=2$ (as it follows from the previous paragraph),
but also for all other values of $n$ it has been possible to check
(up to $n=5$). Since the value of $\lambda_{n}$ rapidly decreases
with the increase of $n$, the same conclusion can be expected to
be valid in the limit of $n\rightarrow \infty$. Table 1
illustrates the dependence of $\epsilon^{(n)}$ on $n$ for four
basic periodic states in the case of $g=1/\sqrt{2}$.

In a Josephson junction array the value of the current in a
junction is also given by Eq. (\ref{f4}), where now
$I_0=(2e/\hbar)J$ ($J$ being the coupling constant of a single
junction). As a consequence, the results of this section are
applicable to a fully frustrated Josephson junction array as well.
The value of the parameter $g$ in the array is also equal to
$g_1$, because its ground state is characterized by exactly the
same values of the variables $\theta_a$, Eq. (\ref{e8}), as in a
fully frustrated wire network just below $T_c$.

On the other hand, for $T\ll T_{c0}$, when
$I_{jk}\propto\theta_{jk}$, the value of $g$ in a fully frustrated
wire network is equal to $g_0=3/5$, as can be found from Eqs.
(\ref{e11}). It can be expected that with the decrease in
temperature the value of $g$ in a network continuously decreases
from $g_1$ to $g_0$. However, the conclusion on the selection of
the state (c) by the magnetic energy is valid not only for
$g_0\leq g\leq g_1$, but in the whole interval $-1<g<1$.

\begin{table}[hb]
\begin{center}
\begin{tabular}{|c||c|c|c|c|}
n & a & c & e & g \\ \hline \hline
2 &  0.5569 &  0.8354 &  -0.5569 & -0.2785 \\ \hline
3 & -0.1704 &  0.6536  & -0.5569 & -0.0966 \\ \hline
4 & -0.1704 &  0.5700  & -0.2227 &  0.1540 \\ \hline
5 & -0.0258 &  0.4977  & -0.3673 &  0.0817
\end{tabular}
\end{center}
\caption{Comparison of dimensionless parameter $\epsilon$
characterizing the magnetic energy of four basic periodic states
in different orders of approximation.}
\end{table}

\section{Conclusion}

The main result of this work is that the superconducting state in
a fully frustrated wire network with the dice lattice geometry
exhibits the same set of degenerate spacial patterns minimizing
the free energy in the two limiting cases when the temperature is
either low compared to the critical temperature $T_c$ (London
limit) or close to $T_c$ (Ginzburg-Landau limit). This conclusion
is quite interesting, given the fact that the system is described
by two rather different models in these two limits. In the London
limit, the amplitude fluctuations of the superconducting order
parameter are negligible, and the corresponding fully frustrated
$XY$ model has been analyzed in~\cite{K01}. In the vicinity of
$T_c$, we used the variational approach pioneered by Abrikosov,
where the spacial variations of the complex order parameter are
constrained to live in the subspace of unstable modes for the
corresponding linearized Ginzburg-Landau equations.

Our interest in this problem had been stimulated by the fact that
for a fully frustrated network with the dice lattice geometry,
this subspace has unusually high degeneracy (as a consequence of
the localized nature of modes) and includes a finite fraction (one
third) of the total number of modes \cite{VMD,Vidal01}.
Remarkably, non linear effects select particular linear
combinations of these spacially localized states which reproduce
precisely the current patterns obtained for the pure $XY$
limit~\cite{K01}.

In the second part of this article, we have compared magnetic
energies of current patterns in different periodic states
minimizing the Ginzburg-Landau functional. The dominant
contribution to this degeneracy lifting mechanism is due to
interactions of current loops which can be associated with second
neighbor plaquettes. It favors the 
periodic pattern in which the triads of positive and negative
vortices have 
three different orientations [state (c) in Fig.~\ref{fig3}]. The
same conclusion is valid also in London limit and in the case of a
Josephson junction array with the same geometry. {It can be hoped
that decoration experiments performed in more equilibrium
conditions than those of Refs. \onlinecite{Pannetier01,Serret02}
may reveal such an ordering.}

This work leaves several open questions. First, it is important to
know if all the states constructed here are stable with respect to local
fluctuations in both order parameter amplitudes and phases.
We have checked numerically that this is indeed the case for the
periodic states shown on Fig.~\ref{fig3},
as well as for configurations with a
single domain wall between two such ordered states. Unfortunately,
a simple (analytical) stability proof holding for the complete class
of degenerate states is not available now.

Second, alternative degeneracy lifting mechanisms should be
analyzed, as has been done for a wire network with {\em kagome}
geometry by Park and Huse~\cite{PH}. These authors have found that
the dominant perturbation to the idealized Ginzburg-Landau
description of a wire network arises from the finite width of
wires, which removes the degeneracy between the two opposite
orientations of the supercurrents in a given loop. This mechanism
can be interpreted in the terms of magnetic field redistribution
between the network plaquettes (see discussion in Ref.
\onlinecite{Meyer}, where it has been named "hidden
incommensurability"), and, accordingly, is effective only when a
network contains non-equivalent plaquettes. Therefore, in the case
of a dice lattice (formed by identical rhombic plaquettes) it
cannot play a prominent role.

Another degeneracy lifting mechanism is related with the free
energy of fluctuations around various free-energy minima
and will be the subject of a separate report.
It is likely to be the dominant one in the vicinity of $T_c$,
but with decrease in temperature becomes less and less important
in comparison with magnetic interactions of currents analyzed
in this work.

\acknowledgments

The work of S.E.K. has been additionally supported by the Program
"Quantum Macrophysics" of the Russian Academy of Sciences and by
the Program "Scientific Schools of the Russian Federation" (grant
No. 1715.2003.2).

\appendix
\section{Condition for the applicability of the mean field approach}

In two-dimensional superconductors the most important fluctuations
are the fluctuations of the order parameter phase
$\varphi=\arg(\Delta)$. When $T_c$ is approached from below, the
free energy of the phase fluctuations in a single wire,
\begin{equation}                                \label{Ap2}
F_{\rm wire}\approx
\frac{J}{2}(\delta\varphi_{j}-\delta\varphi_{k})^2\;,
\end{equation}
can be characterized by the effective coupling constant $J$, which
in the absence of external magnetic field is given by
\begin{equation}                                \label{Ap3}
J(T)=\frac{\alpha(T)\gamma}{L\beta}=\frac{\gamma^2}{L\beta\xi^2(T)}\;.
\end{equation}

The fluctuations are of no importance when \makebox{$J(T)\gg k_B
T$}, which means that the critical region corresponds to
\begin{equation}                                \label{Ap4}
\xi^2(T)\gtrsim \frac{\gamma^2}{L\beta k_BT_{c0}}\;.
\end{equation}
Since the maximal deviation of $T_c(f)$ from $T_{c0}$ is achieved
when $\xi[T_c(f)]\sim L$ \cite{A,PCRV}, the critical region can be
considered as sufficiently narrow if
\begin{equation}                                \label{Ap5}
L^3\ll \frac{\gamma^2}{\beta k_BT_{c0}}\;.
\end{equation}

When the mean free path of electron is much smaller than the thickness
of a wire, the values of the coefficients entering the Ginzburg-Landau
functional for the wire are determined simply by the values of the
analogous coefficients for the material from which the wire is
fabricated. Substitution of
\begin{equation}                                \label{Ap7}
\beta\approx \sigma\beta_{\rm bulk}\,,~~
\gamma\approx \sigma\gamma_{\rm bulk}\,,
\end{equation}
where $\sigma$ is the cross section of a wire, and
\begin{equation}                                \label{Ap6}
\beta_{\rm bulk}=\frac{16\pi^3}{\Phi_{0}^{2}}{\gamma_{\rm
bulk}^{2}\kappa^{2}}\;,
\end{equation}
where $\kappa$ is the ratio of the penetration depth and the
coherence length, into Eq. (\ref{Ap5}) allows one to rewrite this
condition as
\begin{equation}                               \label{Ap8}
L^3\ll \frac{\Lambda_{\rm univ}(T_{c0})\sigma}{\pi\kappa^2} \;,
\end{equation}
where
\begin{equation}                                \label{Ap9}
\Lambda_{\rm univ}(T)=\frac{\Phi_0^2}{16\pi^2 k_B
T}\approx\frac{2\,{{\rm cm}\cdot}{\rm K}}{T}
\end{equation}
is the expression for the universal value of the two-dimensional
penetration length at the temperature of the
Berezinskii-Kosterlitz-Thouless phase transition in a
two-dimensional superconductor \cite{BMO}.  In the right-hand side
of Eq. (\ref{Ap9}) the temperature should be expressed in Kelvin.

In aluminum wire networks fabricated with electron beam
lithography \cite{PCR,PCRV,Abil,Pannetier01,GG,Xiao02}
$T_{c0}\approx 1.2\,$K, and therefore \makebox{$\Lambda_{\rm
univ}(T_{c0})\approx 1.7\,$}cm, whereas (according to the
estimates of Park and Huse \cite{PH}) $\kappa\sim 1$.

\section{Consistency of the equal amplitude hypothesis}
\label{Consistency}

We shall now check that for the class of states described in
section~\ref{Minimization}, in which the amplitudes $|\Delta_{j}|$
are the same for all six-fold coordinated sites $j$, the
derivatives ${\partial F^{(4)}_{{\rm nw}}}/{\partial
|\Delta_{j}|}$ are independent of the sixfold coordinated site
$j$. If this property is satisfied, it is then possible to enforce
the equilibrium condition ${\partial [F^{(2)}_{{\rm
nw}}+F^{(4)}_{{\rm nw}}]}/{\partial |\Delta_{j}|}=0$ with equal
amplitudes $|\Delta_{j}|$ since $F^{(2)}_{{\rm nw}}$ is
proportional to $\sum_{j}|\Delta_{j}|^{2}$ (with a negative
coefficient below the critical temperature of the network).

\begin{figure}[h]
\includegraphics[width=60mm]{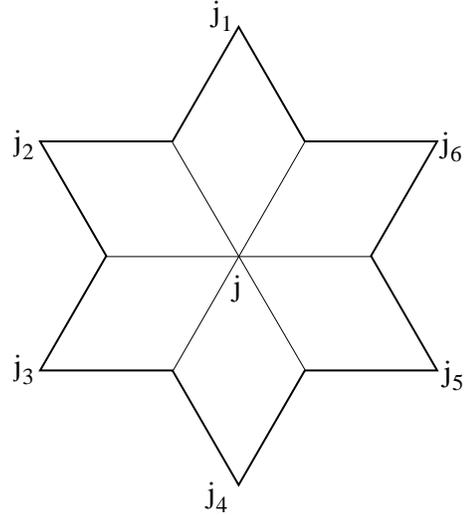}
\caption{\label{figap} The six neighbors of the site $j$
contributing to the sums in Eq. (\ref{derF4}).}
\end{figure}

Derivation of $F^{(4)}_{{\rm nw}}=\sum_{k} F^{(4)}_{{\rm nw}}(k)$
with respect to $|\Delta_{j}|$ gives
\begin{equation}                                \label{derF4}
\frac{\partial F^{(4)}_{{\rm nw}}}{\partial |\Delta_{j}|}=
F_4\Delta^3\left[8\nu_0+4\nu_2\sum_{b=1}^{6}\sin^2\chi_{jj_{b}}-
2\nu_1 \sum_{b=1}^{6}\sin\chi_{j_b j_{b+1}}\right]\;,
\end{equation}
where $j_b$ (with $b=1,\ldots,6$) are the six neighbors of $j$
numbered in positive direction (see Fig.~\ref{figap}), and it is
assumed that $j_{7}\equiv j_{1}$. The contribution from the third
term in Eq. (\ref{ordre4}) vanishes from Eq. (\ref{derF4}) as a
consequence of Eq. (\ref{e3b}), which is valid for any tripod in
any of the states discussed in section~\ref{Minimization}.

It turns out that the first sum in Eq. (\ref{derF4})
is equal to $2$ for any $j$ in any of the considered degenerate
states. On the other hand, summation of Eq. (\ref{e3b}) over the six
tripods containing the given site $j$ allows one to conclude that the
second sum in Eq. (\ref{derF4}) is always equal to zero.
Thus, the expression in the right hand side of
Eq. (\ref{derF4}) does not depend on $j$ for any
configuration with $|\Delta_{j}|=\mbox{const}$ described in
section~\ref{Minimization}.

\begin{widetext} $~$ \end{widetext}

\end{document}